\documentclass[%
 reprint,
 amsmath,amssymb,
 aps,
]{revtex4-2}

\usepackage{graphicx}
\usepackage{dcolumn}
\usepackage{bm}

\begin{document}

\preprint{APS/123-QED}

\title{Mechanism of Conductivity Enhancement of Polymers Employing Microbubble Lithography}

\author{Anand Dev Ranjan\textsuperscript{1}}\email{adr20rs068@iiserkol.ac.in}
\author{Dhananjay Mahapatra\textsuperscript{1}}
\author{Partha Mitra\textsuperscript{1}}
\author{Ayan Banerjee\textsuperscript{1}}\email{ayan@iiserkol.ac.in}

\affiliation{\textsuperscript{1}Department of Physics, IISER Kolkata, Mohanpur, West Bengal, India, 741246}








\begin{abstract}{The pursuit of green methodologies for fabricating optoelectronic devices necessitates the adoption of self-assembly-based strategies to engineer efficient and sustainable platforms. Microbubble lithography (MBL) stands out as a directed self-assembly technique, enabling real-time micropatterning of conductive structures. Notably, this approach achieves significant enhancements in the conductivity of patterned polymers without requiring external dopants. However, the underlying mechanisms driving this enhancement remain poorly understood. In this study, we address this knowledge gap through a combined theoretical and experimental investigation of a binary polymer system. Molecular dynamics simulations and percolation theory reveal structural transformations that underpin improved charge transport. Furthermore, we demonstrate that phase separation at the interfaces of interacting polymers plays a pivotal role in enhancing conductivity. This separation optimizes the conformational states of the polymers, facilitating more efficient charge carrier transport and ultimately leading to higher conductivity. Our findings establish MBL-induced self-assembly as a robust and sustainable technique for fabricating conductive patterns, paving the way for its integration into next-generation optoelectronic devices.}
\end{abstract}

\maketitle

\section{Introduction}
Optoelectronic devices, which form the backbone of modern technologies such as displays, solar cells, and sensors, require precise engineering of their components to achieve optimal efficiency and functionality. Traditional fabrication techniques often involve multiple steps, the use of dopants, and energy-intensive processes, posing challenges in terms of scalability and environmental impact. In this context, self-assembly-based methods present a transformative alternative by leveraging inherent material properties to drive assembly without external additives or complex protocols. In addition, the growing demand for sustainable and efficient fabrication techniques in optoelectronics has also spurred significant interest in self-assembly-based approaches. Importantly, these methods align with global environmental goals by reducing the reliance on hazardous chemicals and resource-intensive processes. Simultaneously, they offer a pathway to simplify manufacturing while improving the performance of functional materials. 

Microbubble lithography (MBL) has emerged as a particularly promising technique within the realm of self-assembly-driven fabrication (\cite{ghosh2020directed}). MBL employs laser-induced microbubbles to pattern materials in real time, enabling precise control over the spatial organization of conductive polymers and other functional materials. This approach eliminates the need for dopants and additional processing steps, positioning it as a green and efficient alternative for fabricating optoelectronic components. Furthermore, MBL offers compatibility with a wide range of materials, making it versatile for various applications in next-generation electronics. A key feature of MBL is its real-time capability, which allows for dynamic adjustments in patterning parameters, thus providing unprecedented control over the self-assembly process. This adaptability makes MBL particularly attractive for applications where precise micro- and nanoscale organization is critical, such as in advanced sensors, flexible electronics, and photonic devices. Moreover, the technique’s ability to operate under ambient conditions further underscores its suitability for large-scale and environmentally friendly manufacturing processes. Despite these advantages, a fundamental understanding of the underlying mechanisms governing MBL-induced self-assembly remains incomplete.

One of the most intriguing aspects of MBL is its ability to enhance the conductivity of patterned polymer systems significantly (\cite{ranjan2023interface}). This enhancement has been observed in various binary systems involving soft oxometalates (SOMs) such as SOM-Pyrrole, SOM-Perylene, and PEDOT:PSS, a widely studied conductive polymer known for its excellent electrical and optical properties (\cite{ghosh2017situ,ranjan2023interface}). PEDOT:PSS consists of two distinct phases: PEDOT, which is conductive, and PSS, which is insulating. The interaction between these phases, along with the ability to manipulate their spatial arrangement, plays a crucial role in determining the overall conductivity of the material. Indeed, our own work has revealed increase in the conductivity of PEDOT:PSS by up to 20x simply by self-assembly and patterning using MBL \cite{ranjan2023interface}. This was qualitatively attributed to phase separation and molecular reorganization -- the former facilitating the segregation of conductive and insulating components, creating distinct pathways for charge transport, while the latter may optimize polymer chain conformations, enhancing charge carrier mobility. However, a quantitative understanding of the interplay between these factors is essential to optimize the performance of MBL-patterned materials.

To address the critical knowledge gaps in the field, we adopt an integrative approach that combines theoretical simulations and experimental validation to elucidate the mechanisms underlying conductivity enhancement in MBL-patterned binary polymer systems. We selected PEDOT:PSS, a widely studied conducting polymer blend, as a model system due to its well-documented electrical, morphological, and structural properties, providing a robust platform to investigate the fundamental processes governing MBL-induced self-assembly. Molecular dynamics simulations are employed to probe the structural rearrangements and charge transport behavior that occur during the self-assembly process. These simulations offer molecular insights into the role of phase separation and polymer alignment in enhancing conductivity, enabling the identification of key parameters driving the observed phenomena. To complement this, percolation theory is applied to analyze the development and connectivity of conductive networks within phase-separated domains. This theoretical framework facilitates deep understanding of the critical thresholds and network formation dynamics that underpin the macroscopic electrical properties of the system. Experimental characterization techniques are then employed to validate and supplement the theoretical findings. X-ray photoelectron spectroscopy (XPS) is utilized to investigate the chemical composition and bonding states, providing detailed information on the molecular interactions influenced by MBL patterning. Simultaneously, atomic force microscopy (AFM) is used to characterize surface morphology and nanoscale features, while scanning electron microscopy (SEM) provides high-resolution imaging of topological changes. These complementary techniques enable a comprehensive analysis of the chemical, morphological, and topological evolutions induced by the MBL process. In addition, by integrating this set of diverse techniques, this study establishes a cohesive framework for understanding the interplay between self-assembly mechanisms and conductivity enhancement in binary polymer systems. The findings not only reveal the intricate processes governing MBL-induced conductivity enhancement but also underscore the potential of this approach for advancing sustainable fabrication techniques. These insights pave the way for scalable and environmentally benign manufacturing strategies in the development of next-generation optoelectronic devices. Finally, the results underscore the potential of self-assembly-based methods in addressing the dual goals of performance enhancement and environmental sustainability in modern optoelectronics.

\section{Materials and Methods}
We use PEDOT:PSS  (Poly(3,4-ethylenedioxythiophene):Polystyrene sulfonate)as our binary polymer system to study the role of self-aseembly process. For this we employed a continuous wave (CW) laser (1064 nm, 100 mW) for the patterning of the semi-conducting polymer employing MBL. The laser was focused onto the PEDOT:PSS dispersion using a 60× objective lens inside an optical tweezers system as already described in detail in literature \cite{ranjan2024biologically,ranjan2024plasmonics,ranjan2023interface,ghosh2020directed}. As we demonstrated in earlier work \cite{ranjan2023interface}, MBL successfully creates linear patterns with uniform dimensions (around 20 µm width, 2 µm height). The measured conductivity of the self-assembled lines was around 800 S/cm, a 5-fold increase compared to pristine PEDOT:PSS (150 S/cm). Here, we used atomic force microscopy (AFM) in tapping mode for analysing the surface topography of the patterned polymers. Scanning electron microscopy (SEM) was employed for capturing high resolution images of the patterned specimens at the nanometer level. The temperature dependent resistance experiments were performed using a four-point geometry inside a cryostat chamber with an attached thermocouple to measure the ambient temperature conditions. 

\section{Results $\&$ Discussion}
We modeled the conductivity enhancement of PEDOT:PSS on the basis of phase separation. PEDOT:PSS is a polymer with a binary component of PEDOT and PSS \cite{chen2002AnnualReviews}. As mentioned earlier, the PEDOT phase is conductive while the PSS phase is insulating. Phase separation in PEDOT:PSS involves the diffusion of PEDOT and PSS molecules to form distinct domains. We used the Monte-Carlo method and the Cahn-Hilliard formalism to describe the energetics and kinetics of this phase separation, which ultimately leads to formation of distinct phases over time. This phase separation process is triggered in our system by the MBL technique, which selectively heats PEDOT regions, inducing localized phase changes and driving self-assembly. Finally, we employed percolation theory over these segregated domains to model the enhanced conductivity. These formalisms are widely used to describe the dynamics of phase separation in binary systems and are particularly effective in capturing the interplay of diffusion and phase boundary evolution. They work on the principle of how PEDOT and PSS domains form and evolve over time as a result of diffusion and local energy minimization \cite{boettinger2002AnnualReviews}. 

\subsection{Framework for Phase Separation}
In binary mixtures like PEDOT:PSS, the co-existence of the two phases (PEDOT-rich and PSS-rich regions) is due to differences in their chemical and physical properties, and is driven by minimizing the free energy of the system. The system tends toward configurations that reduce the overall energy while satisfying thermal fluctuations. Phase separation occurs when a homogeneous mixture of two or more components (e.g., PEDOT and PSS) becomes thermodynamically unstable and separates into distinct regions (phases) with different compositions. In the case of PEDOT:PSS: PEDOT forms a conductive phase, while PSS an insulating phase. The free energy of mixing ($G_{mix}$) dictates whether the system remains mixed or phase separates. The Gibbs free energy of mixing is given by:
\begin{equation}
   G_{mix} = H_{mix} + T \cdot S_{mix} 
\end{equation}
where: $H_{mix}$ is the enthalpy of mixing, representing the energetic interactions between PEDOT and PSS, T is the absolute temperature and $S_{mix}$ is the entropy of mixing, representing the disorder in the system.
We employed the Monte Carlo method to simulate the evolution of the system toward equilibrium by iteratively flipping the phase states. This probabilistic approach allows the system to explore configurations and settle into a lower-energy state. We assumed the interaction energy to be local, involving only nearest neighbors and our system consist of only two possible states (PEDOT-rich or PSS-rich regions). As our binary mixture is composed of two components PEDOT(P) and PSS(S), their respective local concentrations can be described by their local 
densities $\phi_P(x, t)$ and $\phi_P(x, t)$, respectively. As they form a binary mixture, we can assume this local concentration to follow the conservation rule:
\begin{equation}
    \phi_P(x, t) + \phi_P(x, t) = 1
\end{equation}
and therefore, we can describe the whole system using only one concentration $\phi(x, t)$ such that:
\begin{equation}
    \phi_P(x, t) := \phi(x,t) \mbox{ and } \phi_S(x, t) = 1 - \phi(x,t)
\end{equation}
Assuming an initial homogeneous concentration of the binary phases, when a relative perturbation in concentration is created, this results in a flux due to the corresponding changes in the chemical potential of the binary phases. This flux is what gives rise to the phase separation and is given by:
\begin{equation}
    \vec{J} = -M\vec{\nabla}(\mu_P - \mu_S)
\end{equation}
where, M is a mobility coefficient, $\mu_P$ and $\mu_S$ are the chemical potentials of PEDOT $\&$ PSS respectivley. The perturbations in concentration give rise to these potentials which is an effect of the potential energy $\chi[\phi]$ change, owing to this change in concentration fluctuation. Hence, the flux depends on this gradient of potential energy with respect to the concentration:
\begin{equation}
    \mu =: \mu_P - \mu_S = \frac{\delta\chi[\phi]}{\delta \phi} \\
    \implies \vec{J} = -M\vec{\nabla}\frac{\delta\chi[\phi]}{\delta \phi}
\end{equation}
This flux drives the continuous dynamical evolution of this spatial distribution of our binary mixture leading to the Cahn-Hilliard equation \cite{cahn1965}. This equation is a result of the assumption that irrespective of the separation in phases, mass should remain conserved. This results in the fact that the observed changes in the concentration over time $\frac{\partial \phi(x, t)}{\partial t}$ is due to the incoming flux across our region of interest ($\vec{\nabla}\cdot\vec{J}$). The homogeneous heating due to MBL leads to the selective heating of the PEDOT, which eventually leads to this perturbation in the effective concentrations of the binary phases. This leads to the evolution of the spatial distribution of our binary mixture, causing a phase separation given by the Cahn-Hilliard equation \cite{kim2016wiley}. The observed phase separation in the binary polymer system during MBL-induced self-assembly as modeled using the Cahn-Hilliard equation can be given by :
\begin{equation}
    \frac{\partial \phi(\vec{r},t)}{\partial t} = -\vec{\nabla}\cdot\vec{J} = -\vec{\nabla}\cdot(-M\vec{\nabla}\mu)  
\end{equation}
\begin{equation}
   \frac{\partial \phi(\vec{r},t)}{\partial t} = M \nabla^2 \mu = \vec{\nabla}\cdot M\vec{\nabla}\frac{\delta\chi[\phi]}{\delta c}
\end{equation}

The chemical potential is derived from the system’s free energy, $\mu = \frac{\delta \chi[\phi]}{\delta \phi}$, which includes both bulk and interfacial contributions. This free energy functional of the system which describes the thermodynamic driving forces for phase separation can be represented using the concentration field $\phi(\vec{r},t)$ as \cite{cahn1958}:
\begin{equation}
    \chi[\phi] = \int[\frac{A}{2}\phi(\vec{r},t)^2(1-\phi(\vec{r},t)^2) + \frac{B}{2}(\nabla \phi(\vec{r},t))^2]dV 
\end{equation}
where, A controls the phase separation strength, and B relates the interfacial energy. Simulations were performed to solve this equation numerically, capturing the time evolution of $\phi(r, t)$. 

\subsubsection{Simulated Phase Separation and Conductive Pathway Formation}
\begin{figure}[h!]
\begin{center}
\includegraphics[width=8.5cm,height=1.75cm]{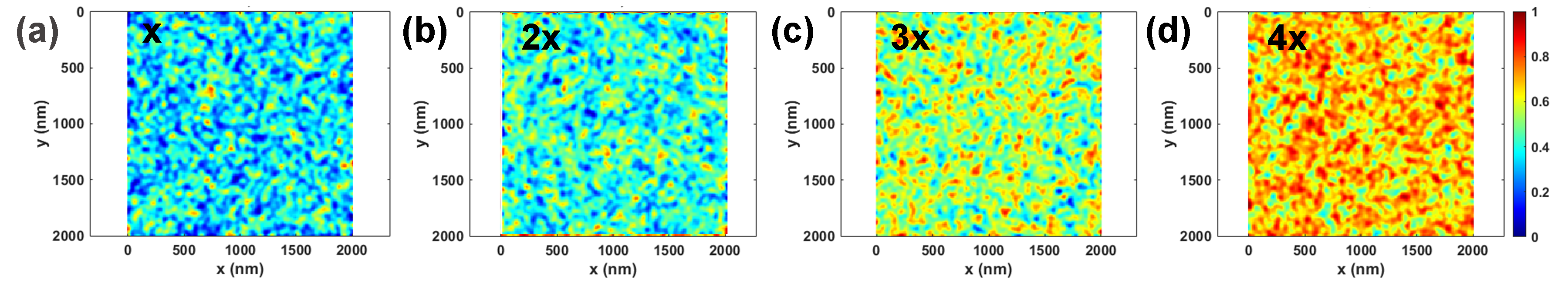}
\end{center}
\caption{(a-d) Evolution of PEDOT phase with a given initial concentrations (x). As we increase the concentration of PEDOT in multiples of this initial concentration (x, 2x, 3x, 4x) the the phase boundaries also increase resulting in larger domains of PEDOT} 
\label{simu}
\end{figure}

These simulations reveal the formation of distinct domains corresponding to the phase-separated regions as shown in Fig.~\ref{simu}(a-d). Each system begins with a randomly distributed initial concentration (x). As this concentration increases in multiples of the initial concentration (x, 2x, 3x $\&$ 4x), phase boundaries expand, resulting in a higher phase separation of the PEDOT layer (reddish boundaries). Consequently, the domain size increases, as observed from Fig.~\ref{simu}(a) to Fig.~\ref{simu}(d). At lower PEDOT concentrations, the phase-separated domains remain small and isolated, restricting the percolation pathways necessary for efficient charge transport. However, as the concentration increases, larger and more interconnected PEDOT-rich domains emerge, facilitating improved conductivity in the PEDOT:PSS system. This behavior highlights a direct correlation between phase separation and electrical conductivity, which can be further analyzed through percolation theory \cite{kirkpatrick1973RevModernPh}.

Figure.~\ref{simu} provides a visual representation of this phase separation process. In (a), at the initial concentration ( x ), PEDOT-rich domains (reddish regions) are sparse and remain largely disconnected, indicating limited phase separation. As the concentration doubles in (b) ( 2x ), the PEDOT domains begin to expand, though they still remain relatively isolated. Further increasing the concentration to (c) ( 3x ) results in more pronounced phase separation, with larger and more interconnected PEDOT-rich regions forming. Finally, at (d) ( 4x ), the PEDOT domains merge into a nearly continuous network, significantly enhancing connectivity. The accompanying color scale represents the local concentration of PEDOT, with blue indicating low concentrations and red signifying high concentrations. This visualization demonstrates how increasing the PEDOT concentration promotes the formation of a conductive percolative network, which is essential for efficient charge transport in the PEDOT:PSS system.

\begin{figure}[h!]
\begin{center}
\includegraphics[width=8.5cm,height=3.25cm]{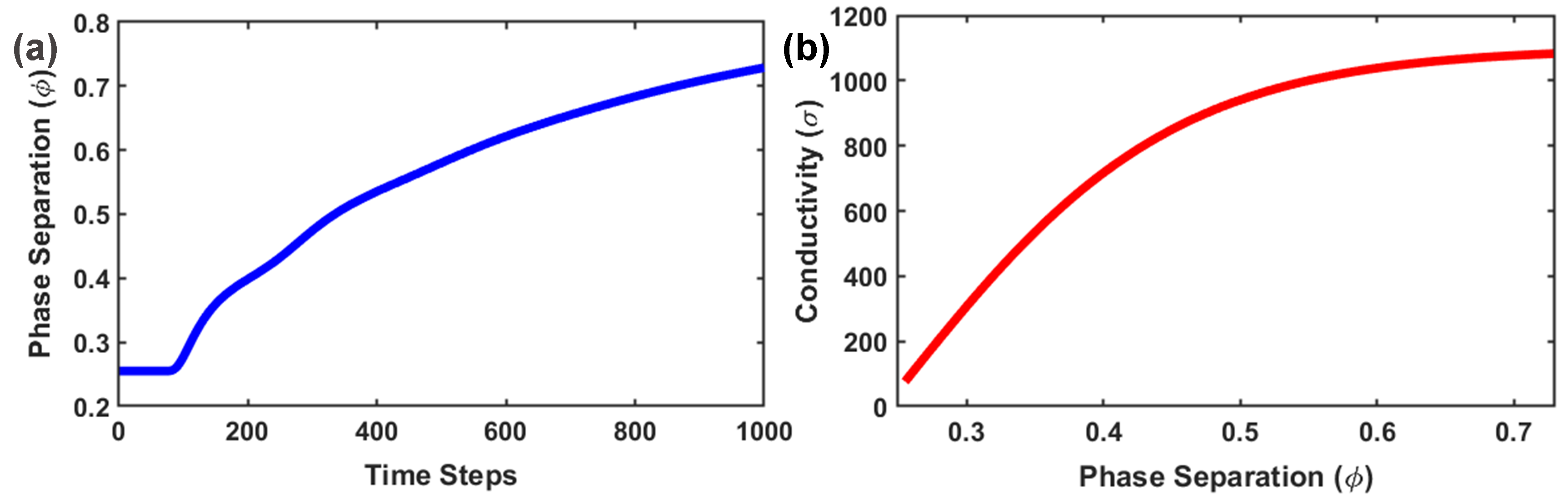}
\end{center}
\caption{Evolution of phase separation with time (b) Conductivity of the binary polymer system with changing phase separation.}
\label{simu-cond}
\end{figure}

The Cahn-Hilliard equation describes the evolution of the local PEDOT concentration, $\phi(r, t)$, capturing the phase separation process. As the simulations reveal (Fig.~\ref{simu}(a–d)), phase separation leads to the formation of PEDOT-rich regions (high $\phi$) and PSS-rich regions (low $\phi$). This evolving domain structure can be modeled statistically as a percolative network, where conductive (PEDOT) and insulating (PSS) regions form a randomly distributed system \cite{ali2022pedot}. A key concept in percolation theory is the percolation threshold, $\phi_c$, which defines the critical concentration at which a continuous conductive network emerges. Since PEDOT is the only conducting phase in this system, $\phi_c$ represents the minimum PEDOT concentration required for charge transport across the material. Below $\phi_c$, PEDOT domains are isolated, and overall conductivity remains negligible. Above $\phi_c$, the system transitions to a percolative regime, where interconnected PEDOT-rich domains establish a conductive network, leading to a significant increase in electrical conductivity. Regions where $\phi(r,t) > \phi_c$ contribute to charge transport, while areas with $\phi(r,t) < \phi_c$ act as insulating barriers \cite{huang2020Sage}. Hence, as the average concentration of conductive regions increases and approaches our critical threshold, the material transitions from low to high conductivity (viz: $\sigma$ is a function of $\phi-\phi_c$). We capture this behaviour of conductivity by incorporating a smooth, sigmoidal increase that mimics percolation, where isolated conductive islands gradually connect to form a continuous pathway \cite{Anderson1984PRB,rahaman2017sigmoidal,jeong2021sigmoidal}.
\begin{equation}
   \sigma(\phi) \propto f(\phi-\phi_c) \implies \sigma(\phi) = \sigma_0 + \frac{\sigma_{max} - \sigma_0}{e^{\beta(\phi-\phi_c)} - 1} \times (e^{\beta(\phi-\phi_c)} + 1)
\end{equation}

where, $\beta$ controls the steepness of transition, $\sigma_0$ $\&$ $\sigma_{max}$ are the base and maximum conductivity of the phases. Since, MBL introduces a abrupt sharper phase owing to its laser assisted self-assembly, therefore we also introduced the additional exponential term to account for this abrupt phase change. Additionally, these also account for saturation of the conductivity as the phase separation is maximised. This saturation mimics the limiting behavior of conductivity once the conductive network is fully established as also observed in multiple polymer systems \cite{shklovskii2013Springer,taherian2016saturation}. Together, these mechanisms provide a technical yet intuitive description of how the microstructural evolution of the material directly influences its macroscopic electrical properties \cite{zhang2024saturation}. In our study, we specifically simulate the phase separation induced by MBL, where laser-induced heating selectively modifies the PEDOT domains. This targeted heating perturbs the local concentration of PEDOT and PSS, leading to phase separation as governed by the Cahn-Hilliard equation \cite{kim2016wiley}. To solve the coupled equations governing this phase evolution, we applied boundary conditions of $\phi(r,t) \rightarrow 1$ for PEDOT domains and $\phi(r,t) \rightarrow 0$ for PSS phases. These boundary conditions ensure that phase separation progresses in a controlled manner, allowing PEDOT-rich regions to coalesce and form conductive pathways. The homogeneous heating due to MBL leads to the selective heating of PEDOT, which eventually results in a perturbation in the effective concentrations of the binary phases. This drives the evolution of the spatial distribution of the binary mixture, causing phase separation, as shown in Fig.~\ref{simu-cond}(a). Initially, phase separation increases rapidly due to thermodynamic instabilities, followed by a slower growth as domains stabilize. As the phase separation progresses from an initial homogeneous mixture to a structured network (Fig.~\ref{simu-cond}(a)), the conductivity of the domains increases accordingly (Fig.~\ref{simu-cond}(b)). This evolution, driven by MBL-induced self-assembly, results in the formation of percolative networks that significantly enhance electrical conductivity. Finally, the saturation of conductivity at higher phase separation indicates that most PEDOT domains become interconnected, resulting in a percolative network capable of efficient charge transport. 

Together these simulations provide insights into the structural evolution of the binary polymer system during MBL. Initially, the system exhibited a homogeneous mixture of the two polymer components. As MBL was performed, energy introduced by the microbubble triggered phase separation due to thermodynamic instabilities. Over time, the domains coarsened, forming interconnected networks that align with percolation theory predictions. Notably, the phase separation not only improved charge carrier mobility but also optimized the conformational states of the polymers, reducing barriers to charge transport. This network formation is crucial for enabling continuous conduction across the patterned region. Guided by these theoretical insights, we next investigate the experimental validation of this phase separation-driven conductivity enhancement.

\subsection{Experimental Validation of Structural Changes
}

To experimentally validate the role of phase separation in enhancing the conductivity of PEDOT:PSS, we patterned PEDOT:PSS films at increasing laser powers. As predicted in the literature, MBL-based self-assembly induces phase separation between PEDOT and PSS chains. By systematically tuning the laser power, we can control the extent of this phase separation. Since MBL selectively heats PEDOT due to its enhanced absorption at the laser wavelength \cite{ranjan2023interface}, higher laser powers introduce more thermal energy into the system. This increased energy accelerates polymer chain mobility, facilitating phase separation by promoting PEDOT aggregation and PSS expulsion. As a result, the degree of phase separation should vary proportionally with laser power. Changing phase separation means changing grain boundaries of the polymer system. Hence, to visualize and quantify these morphological changes induced by phase separation, we performed surface morphology analysis using scanning electron microscopy (SEM) and validated these observations with atomic force microscopy (AFM). Additionally, X-ray photoelectron spectroscopy (XPS) and electrical measurements, including I-V characteristics and resistance vs. temperature (R-T) analysis, were conducted to further correlate the extent of phase separation with changes in charge transport behavior.

\subsubsection{Surface Morphology}
\begin{figure}[h!]
\begin{center}
\includegraphics[width=8.5cm,height=2.5cm]{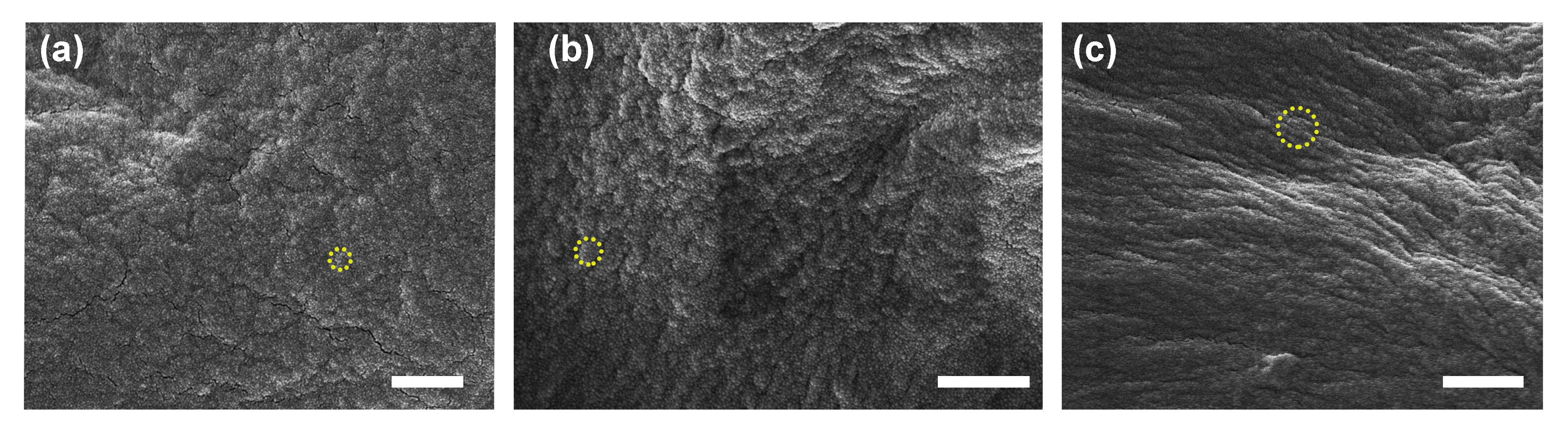}
\end{center}
\caption{(a-c) Surface topography of patterns of PEDOT:PSS formed at varying laser power of 20mW, 30mW $\&$ 40mW respetively. Clearly the grain size increases with increasing laser power with maximum grain size at 40mW and minimum grain size at 20mW.}\label{sem}
\end{figure}

SEM analysis was conducted to examine the effect of laser power on the morphology of self-assembled PEDOT:PSS patterns. Thus, PEDOT-PSS was patterned at 3 different laser powers - 20, 30 $\&$ 40 mW - the SEM images of the corresponding patterns being shown in Figs.~\ref{sem} (a)--(c), respectively. The figures clearly reveal that increasing the laser power leads to significant enlargement of the PEDOT phase domains within the PSS matrix. At the lowest laser power (20 mw), the PEDOT phase appears as smaller, with hard to visualise distinct domains dispersed within the PSS matrix. These domains are relatively uniform in size, suggesting limited thermal forces during the self-assembly process. As the laser power is increased, the PEDOT domains grow in size, becoming more prominent and less uniformly distributed. This growth indicates that higher laser powers facilitate enhanced diffusion and coalescence of PEDOT within the PSS matrix, likely driven by increased local heating due to the higher power of the laser, and optical gradient forces generated by the focused laser light. Quantitative analysis of the SEM images confirms these observations. Thus, at a laser power of 20 mW, the average domain size is approximately 91 ± 26 nm, whereas at 40 mW, the average domain size increases to 202 ± 17 nm. This systematic enlargement suggests that the MBL-induced thermal effects play a critical role in the self-assembly dynamics. This increase in the domain size further establishes our hypothesis that MBL enhances the phase separation between the polymer system. To further establish these observations, AFM analysis was conducted on the similar samples. 

\subsubsection{Surface Topography}
\begin{figure}[h!]
\begin{center}
\includegraphics[width=8.5cm,height=2.5cm]{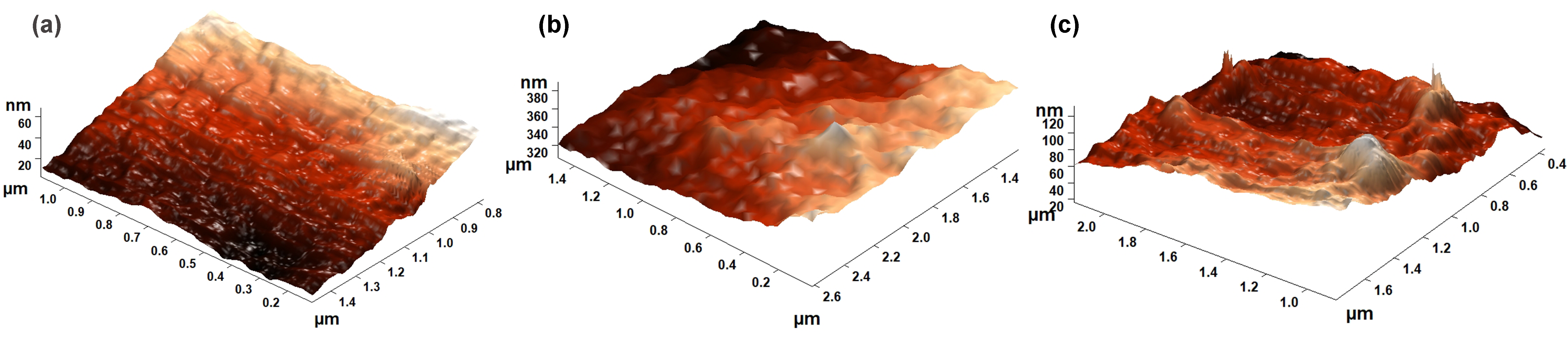}
\end{center}
\caption{(a-c) Surface topography of patterns of PEDOT:PSS formed at varying laser power of 20mW, 30mW $\&$ 40mW respectively. Clearly the grain size increases with increasing laser power with maximum grain size at 40mW and minimum grain size at 20mW.}\label{afm-sem}
\end{figure}

AFM was employed to validate the morphological observations from the SEM data, and to gain additional insights into the topographical properties of the self-assembled patterns. The results are shown in Fig.~\ref{afm-sem}(a-c), revealing a phase-separated morphology with distinct domains, consistent with the patterns predicted by the Cahn-Hilliard model. Height profiles showed that the phase-separated regions exhibited a topographical contrast, further confirming the formation of two distinct phases.  

In addition, the AFM image of the PEDOT:PSS film displayed distinct nanoscale grains indicative of MBL-induced self-assembly.  The grain size distribution, as measured using the software {\it ImageJ}, ranged from 78 nm ± 29 nm to 215 ± 25 nm. This result is consistent with prior studies, which report that laser power influences the degree of polymer chain aggregation and grain growth. Importantly, this roughness correlates with enhanced phase separation, which is expected to contribute to improved conductive pathways. 

The complementary nature of the SEM and AFM results strengthens the conclusion that laser power significantly influences the self-assembly dynamics of PEDOT:PSS. The AFM analysis also provides additional topographical information, confirming the uniformity and three-dimensional characteristics of the enlarged PEDOT domains at higher laser powers. These results provide key insights into the tunability of PEDOT:PSS self-assembly using laser power, highlighting the potential for precise control over domain size and morphology in applications requiring tailored conductive patterns.

\subsubsection{Chemical Composition}
To complement the structural observations, XPS was employed to analyze the chemical environment of the patterned polymers. XPS revealed shifts in the binding energies of carbon and oxygen functional groups, indicative of chemical transformations induced by MBL. Thus, there appeared a relative enrichment of PEDOT at the surface of the domains, especially at higher laser powers. This enrichment is consistent with the observed phase separation and suggests that laser-induced thermal effects promote the migration and accumulation of PEDOT at the phase boundaries. Furthermore, the increased PEDOT concentration correlates with enhanced conductivity, as PEDOT is the primary conductive component in the system.

\begin{figure}[h!]
\begin{center}
\includegraphics[width=9cm,height=2.25cm]{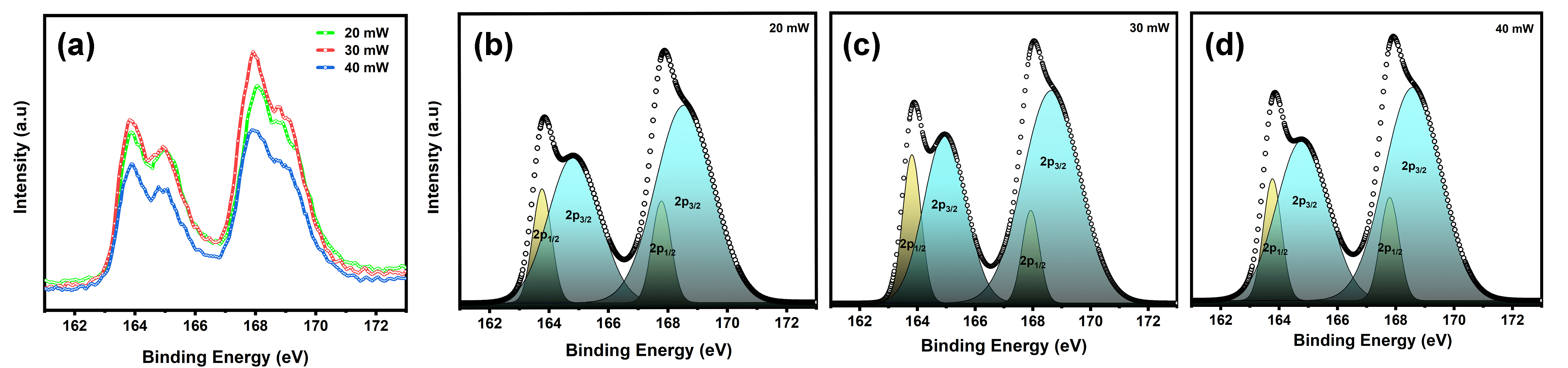}
\end{center}
\caption{ (a) XPS of PEDOT:PSS self-assembled at various laser powers (b-d) S2p XPS spectra of the samples clearly reveal the increase in PEDOT phase compared to the PSS phase.}\label{xps}
\end{figure}

Fig.~\ref{xps}(a-b) illustrates the XPS spectra for S (2p) of pristine and self-assembled PEDOT:PSS. The S (2p) bands are observed at the binding energy between 162 and 166 eV that correspond to the sulfur atoms of PEDOT, and at the 168 and 170 eV that correspond to the sulfur signal from PSS. The ratio of PEDOT and PSS was calculated to 1:1.36 for the pristine film by calculating the area under the curve. For the self-assembled case, S (2p) from PEDOT shows significantly higher percentage of PEDOT as clearly seen from the spectra itself (1:0.73).

Our XPS spectra show an increase in signals associated with pi-pi stacking interactions, suggesting enhanced alignment of conjugated polymer backbones. This chemical reorganization facilitates more efficient charge-carrier mobility, further supporting the role of phase separation in optimizing the material's electronic properties. The combined insights from SEM, AFM, and XPS confirm that MBL-induced self-assembly effectively enhances phase separation in PEDOT:PSS, resulting in a morphology that is likely to promote superior charge transport characteristics.



\subsection{Correlation Between Phase Separation and Conductivity} 
\begin{figure}[h!]
\begin{center}
\includegraphics[width=8cm,height=2.3cm]{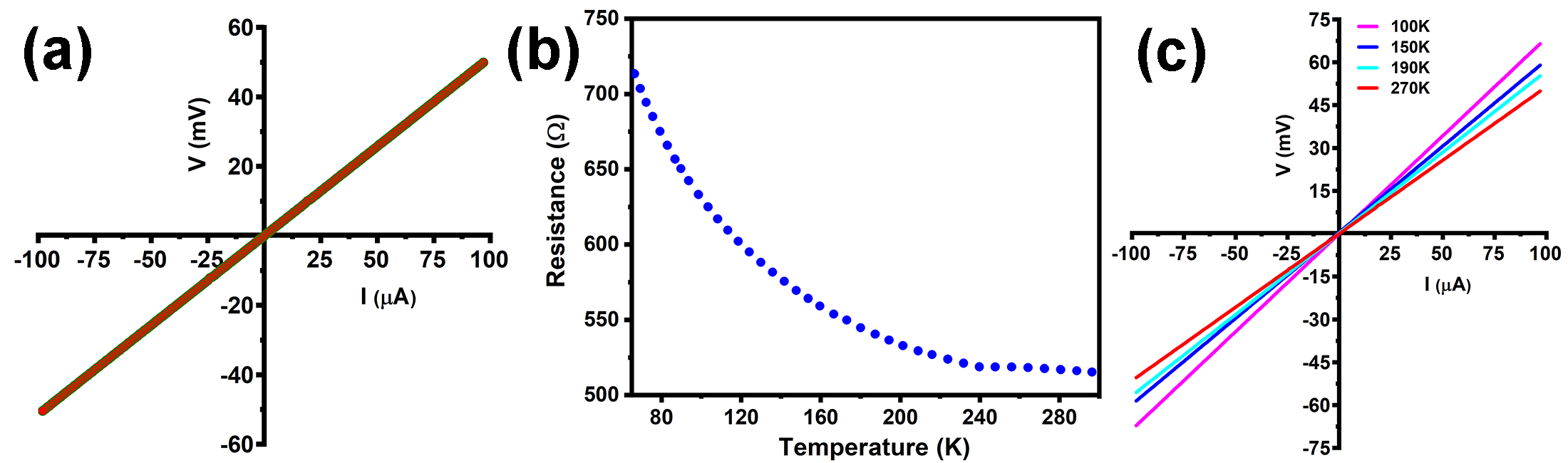}
\end{center}
\caption{ (a) IV characteristics of the bi-polymer system self-assembled employing MBL (b) The change in the resistance of the pattern with respect to temperature (c) The IV graph of pattern at various temperature shows a linear behaviour.}
\label{condt-RT}
\end{figure}

The SEM, AFM, and XPS analyses performed at increasing laser powers confirm that MBL induces phase separation, resulting in increased PEDOT aggregation and a higher charge carrier density in the patterned polymer system. However, while these techniques reveal structural and compositional changes, they do not provide direct insights into the transport characteristics of these charge carriers. Since conductivity in disordered polymers is often thermally activated at higher temperatures -- implying the absence of a well-percolated conductive network -- it is essential to assess the transport properties to establish the impact of phase separation on conductivity. To address this, we performed current-voltage (I-V) and resistance-temperature (R-T) studies on the polymer pattern formed at the highest power (40 mW), where phase separation was most pronounced. In addition, this pattern shows significantly larger PEDOT-rich domains and improved phase separation, as confirmed by SEM and AFM, which are expected to facilitate enhanced charge transport through well-connected conductive pathways. We first performed current-voltage (IV) measurements at room temperature on these samples. The obtained I-V curve (Fig.~\ref{condt-RT}(a)) exhibits a linear relationship, confirming ohmic conduction. This suggests that the self-assembly process results in well-connected PEDOT-rich pathways, minimising interfacial barriers and ensuring efficient charge transport. The absence of non-linearity in the I-V characteristic indicates that the charge carriers experience minimal tunneling, supporting the hypothesis of enhanced conductivity as a result of optimized phase separation. 

To further investigate the transport mechanism, we measured the resistance as a function of temperature (R-T) from 100K to 300K (Fig.~\ref{condt-RT}(b)). The resistance exhibits a monotonic decrease with increasing temperature, consistent with thermally activated charge transport in conducting polymers. However, a notable feature of the R-T curve is the saturation of resistance at higher temperature ($\sim$ 240K). This deviation from purely thermally activated behaviour suggests that at higher temperature, the charge mobility is no longer significantly limited by the energy barriers, reinforcing the idea that the self-assembly process has facilitated efficient charge percolation paths. In contrast, conventional PEDOT:PSS films typically show a continuous trend without saturation, further emphasising the impact of the MBL-assisted assembly on conductivity enhancement \cite{seifi2022RT}.

To validate the trend observed in R-T analysis, we performed I-V measurements at multiple temperatures as shown in Fig.~\ref{condt-RT}(c). The linearity of I-V curves across various temperatures confirms that conduction remains ohmic and is not dominated by charge trapping or variable-range hopping (VRH). In typical disordered polymer systems, lower temperature often leads to non-ohmic behaviour due to charge localization effects; however, self-assembled PEDOT:PSS patterns maintain ohmic conduction even at low temperatures. This further supports the hypothesis that the MBL-assisted assembly leads to an optimised phase separation, where PEDOT-rich regions provide continuous conductive pathways with reduced localisation. 

The combined electrical characterization -- which includes ohmic conduction at all temperatures, resistance saturation at higher temperatures, and suppressed localization effects at low temperatures --strongly suggests that phase separation plays a crucial role in determining the overall conductivity of the self-assembled PEDOT:PSS system. The MBL-induced self-assembly promotes a percolative conductive network of PEDOT-rich domains, reducing charge carrier scattering and enhancing the overall conductivity. These observations demonstrate that MBL-induced self-assembly at optimized laser power enhances the morphological and chemical properties of PEDOT:PSS, resulting in improved charge transport characteristics.  The self-assembly approach thus provides a highly tunable and efficient method to improve the electronic properties of PEDOT:PSS for flexible electronics and sensor applications.

\section{Conclusions}
This study explores the physics behind the enhancement of conductivity of PEDOT:PSS patterns fabricated using MBL \cite{ranjan2023interface}. The ability to achieve such a substantial enhancement in conductivity is attributed to a complex interplay between phase separation, densification of PEDOT domains, and improved percolation pathways, which were thoroughly investigated using both simulations and experiments. MD simulations provided a theoretical framework to understand the mechanism behind the conductivity enhancement. This successfully modeled the phase separation process, predicting the redistribution of PSS and densification of PEDOT domains, which leads to the formation of conductive pathways. As the polymer system undergoes these transformations, the Cahn-Hilliard formalism accurately captured the kinetics of phase separation, showing the formation and evolution of PEDOT-rich and PSS-rich regions. The simulations further indicated that phase separation reaches a critical percolation threshold, where the PEDOT-rich domains form continuous conductive pathways. These percolating pathways reduce the resistance by facilitating efficient charge transport, consistent with percolation theory. The simulations also highlighted how the molecular alignment and densification of PEDOT domains contribute to the observed conductivity enhancement, demonstrating that the laser treatment significantly alters the structural and morphological characteristics of the polymer system.

Molecular dynamics simulations provide crucial insights into the structural transformations that occur during MBL-induced self-assembly. The simulations indicate that the interaction of laser energy with the binary polymer system promotes phase separation, leading to the formation of distinct conductive pathways. Percolation theory analysis reveals that these pathways achieve a critical density, facilitating efficient charge transport across the patterned structure. The significant conductivity enhancement achieved through phase separation, densification, and molecular alignment establishes this method as a promising tool for next-generation optoelectronic applications. Future work will focus on optimizing laser parameters and exploring new applications.
The Cahn-Hilliard formalism successfully captured the kinetics of phase separation, while the experimental techniques (AFM, SEM, and XPS) provided complementary evidence of the structural and chemical changes. Together, these results demonstrate that MBL-induced self-assembly drives phase separation and molecular reorganization, leading to significant conductivity enhancements.

The IV characteristics and temperature-dependent resistance data collectively support a strong correlation between conductivity enhancement and reduced phase separation in the laser-treated self-assembled PEDOT:PSS patterns. The linear IV response at room temperature indicates that the laser treatment improved the charge transport by facilitating better connectivity between the PEDOT domains, which is a key factor for reducing the negative impact of phase separation. The weaker temperature dependence in the resistance measurements further confirms that the material has become less prone to the insulating effects of PSS, as phase-separated regions are likely minimized.

In more detail, the observed ohmic behavior and reduced temperature dependence point to the laser treatment leading to a more uniform distribution of PEDOT within the self-assembled patterns. This would allow for improved percolation pathways, meaning that charge carriers can travel more freely across the film without encountering significant barriers or isolated PEDOT domains. By reducing phase separation, the treatment appears to have promoted a more stable conductive network that performs consistently across a range of temperatures.

The combined results from SEM, AFM, and XPS provide a comprehensive understanding of the role of phase separation in conductivity enhancement. The enlargement of PEDOT domains, increased surface roughness, and enrichment of PEDOT at the domain boundaries collectively contribute to the formation of more efficient conductive pathways. These findings highlight the significance of controlled phase separation in optimizing the performance of PEDOT:PSS-based conductive materials.

These results demonstrate that phase separation, driven by laser-induced effects, is a critical factor in enhancing the conductivity of PEDOT:PSS. The integration of SEM, AFM, and XPS analyses offers a robust framework for understanding the interplay between morphology, composition, and conductivity in this system.The findings contribute to the understanding of how local modifications, such as laser treatments, can effectively manipulate phase separation in conductive polymer systems and optimize performance for various electronic applications.

\section*{Acknowledgments}
This is a short text to acknowledge the contributions of specific colleagues, institutions, or agencies that aided the efforts of the authors.

\bibliography{test}

\end{document}